\documentclass[aps,prd,preprint,nofootinbib,longbibliography]{revtex4-1}

\usepackage{amsmath}
\usepackage{amssymb} 
\usepackage[cm]{fullpage}
\usepackage{graphicx} 
\usepackage{siunitx}
\usepackage{tabularx} 
\usepackage[colorlinks,linkcolor=blue,citecolor=blue,urlcolor=blue]{hyperref}
\usepackage[table,xcdraw]{xcolor}
\usepackage[utf8]{inputenc}

\DeclareSIUnit\parsec{pc}

\newcommand{\beq}{\begin{equation}}
\newcommand{\eeq}{\end{equation}}
\newcommand{\beqx}{\begin{equation*}}
\newcommand{\eeqx}{\end{equation*}}
\newcommand{\beqa}{\begin{eqnarray}}
\newcommand{\eeqa}{\end{eqnarray}}
\newcommand{\beqax}{\begin{eqnarray*}}
\newcommand{\eeqax}{\end{eqnarray*}}

\newcommand{\recfrac}[2]{\frac{#2}{#1}}

\newcommand{\eps}[1]{\epsilon_{#1}}
\newcommand{\del}[1]{\delta_{#1}}
\newcommand{\Del}[1]{\Delta_{#1}}
\newcommand{\zet}[1]{\zeta_{#1}}

\newcommand{\epsd}[1]{\dot{\epsilon}_{#1}}
\newcommand{\deld}[1]{\dot{\delta}_{#1}}
\newcommand{\zetd}[1]{\dot{\zeta}_{#1}}
\newcommand{\epsdd}[1]{\ddot{\epsilon}_{#1}}
\newcommand{\deldd}[1]{\ddot{\delta}_{#1}}
\newcommand{\zetdd}[1]{\ddot{\zeta}_{#1}}

\newcommand{\rpar}[1]{\left(#1\right)}
\newcommand{\spar}[1]{\left[#1\right]}

\newcommand{\covd}[1]{\rpar{\nabla #1}}
\newcommand{\cbbox}[1]{\rpar{\Box #1}}
\newcommand{\bd}{{\rm d}}

\newcommand{\eqtxt}[1]{\mathrel{\overset{\makebox[0pt]{\mbox{\normalfont\tiny\sffamily #1}}}{=}}}

\begin{document}

\title{Higgs Inflation with a Gauss-Bonnet term in the Jordan Frame}
\date{\today}
\author{Carsten van de Bruck}
\author{Chris Longden}
\affiliation{Consortium for Fundamental Physics, School of Mathematics and Statistics, University of Sheffield, Hounsfield Road, Sheffield S3 7RH, United Kingdom}

\begin{abstract}
We consider an extension of Higgs inflation in which the Higgs field is coupled to the Gauss-Bonnet term. Working solely in the Jordan frame, we firstly recover the standard predictions of Higgs inflation without a Gauss-Bonnet term. We then calculate the power spectra for scalar and tensor perturbations in the presence of a coupling to a Gauss-Bonnet term. We show that generically the predictions of Higgs inflation are robust and the contributions to the power spectra coming from the Gauss-Bonnet term are negligible. We find, however, that the end of inflation can be strongly modified and that we hence expect the details of (p)reheating to be significantly altered, leading to some concerns over the feasibility of the model which require further investigations.
\end{abstract}

\maketitle

\tableofcontents

\section{Introduction}

While the idea of inflation, an accelerating expansion in the very early universe, has been very successful in circumventing problems with the standard big bang cosmology, as well as accounting for the near-scale-invariant spectrum of primordial density perturbations observed now to a high level of precision by Planck \cite{Planck2015Inflation}, there is still no conclusive solution to the problem of how to embed inflation into a particle physics motivated framework. The most common approach to this problem has been to couple gravity to a scalar field, the inflaton, in the context of a theory beyond the standard model. The inflaton in such models may be based on scalar degrees of freedom that arise in theories such as supergravity, braneworld scenarios, or even modifications of general relativity, such as is the case with Starobinsky inflation \cite{Starobinsky:1980te}. 

More recently, the idea that the standard model Higgs boson may be the inflaton \cite{Bezrukov:2007ep,Bezrukov:2008ej,Barvinsky:2008ia,GarciaBellido:2008ab} has attracted considerable interest. Especially with the discovery and analysis of the properties of the Higgs at the Large Hadron Collider \cite{HiggsDiscovery1, HiggsDiscovery2}, this particular model now has considerably less uncertainty in its parameters than those which rely on embeddings in poorly constrained extensions of the standard model (see \cite{Bezrukov:2013fka} for a review). Furthermore, it agrees with data from CMB experiments surprisingly well. The key ingredient which makes Higgs inflation work is a non-minimal coupling between the Higgs field and the gravitational sector\footnote{A different type of interaction than the one we consider has been studied in \cite{Germani:2010gm}.}, which flattens the Einstein-frame potential in the large-field regime, allowing the slow-roll conditions for inflation to be realised. It is well known that at first order in slow-roll, this model predicts a spectral index which is consistent with data, as well as a low enough tensor-to-scalar ratio to be comfortably below the experimental limits on primordial gravitational waves. 
 
Despite its status as one of the most minimalistic, yet experimentally compatible, models of inflation, it does however have some issues. The required magnitude of the non-minimal coupling has been criticised as unnatural, in an effective field theory sense, and there are concerns over whether the \mbox{(meta-)stability} of the vacuum in Higgs inflation scenarios is sufficient (see \cite{Bezrukov:2014ipa} for a discussion). Furthermore, while the presence of the non-minimal coupling to gravity in Higgs inflation is in itself well-motivated by consideration of the renormalisation of a scalar field in curved space \cite{Birrell1982}, it is quite feasibly not the end of the story when it comes to additional interactions that one would expect to be present when the model is embedded in some UV-complete theory \cite{Bezrukov:2010jz}. One particularly simple, yet potentially interesting modification of this type would be to consider couplings to quadratic and/or higher-order curvature scalars, the simplest manifestation of which would be the Gauss-Bonnet combination $R^2 - 4 R^{\mu \nu} R_{\mu \nu} + R^{\rho \mu \sigma \nu} R_{\rho \mu \sigma \nu}$, and that is the aim of this paper. Such motivations for the addition of the Gauss-Bonnet term are complemented by considerations from string theory where particular couplings between the Gauss-Bonnet term and scalar fields have been found \cite{Kawai:1997mf, Kawai:1998ab, Kawai:1999pw}. 

The inflationary behaviour of a generic scalar field coupled to just the Gauss-Bonnet term has been studied in isolation \cite{Satoh:2007gn,Satoh:2008ck,Guo:2009uk,Guo:2010jr,Nozari:2013wua}, but this is equivalent to adding the Gauss-Bonnet term in the Einstein frame version of a theory\footnote{More general couplings to any function of the Gauss-Bonnet combination have also been considered in \cite{DeLaurentis:2015fea}}. Theories with modifications such as non-minimal couplings to gravity such as Higgs inflation, however, are often formulated in the Jordan Frame, and as such additional couplings such as to the Gauss-Bonnet combination may be more meaningful when implemented directly in the Jordan frame.  We are hence motivated to study inflation in the context of a scalar field with both a Gauss-Bonnet coupling and a non-minimal coupling to the Ricci scalar in the action. 

The paper is organised as follows. In the next section, we present the model. In section III, we study standard Higgs inflation (without Gauss--Bonnet term), but perform the calculation in the Jordan frame, instead of the Einstein frame, as it is usually done in the literature. This is to set the scene for the calculation when we include the Gauss--Bonnet term but it also acts as a consistency check of our approach. Not surprisingly, we find that the Jordan frame calculation agrees with the Einstein frame one. In Section IV we include then the Gauss-Bonnet term and find expressions for the scalar perturbations and the tensor perturbations. In Section V we discuss the results and the implications for the end of inflation. Our conclusions can be found in section VI. 

\section{The Model}

We consider a scalar-tensor theory specified by the action,

\beq \label{eq:gbhaction}
S = \int \bd^d x \sqrt{-g} \left[ \frac{1}{2}F(h) R - \frac{1}{2} \omega (\partial h)^2 - V(h) - \frac{1}{2} G(h) E \right] \, ,
\eeq
where $E=R^2 - 4 R^{\mu \nu} R_{\mu \nu} + R^{\rho \mu \sigma \nu} R_{\rho \mu \sigma \nu}$ represents the Gauss-Bonnet combination, and the potential $V(h)$, non-minimal coupling function $F(h)$ and Gauss-Bonnet coupling function $G(h)$ are unspecified functions of the scalar field $h$. If we interpret $h$ as the unitary-gauge Higgs field, this action is a generalisation of the standard Higgs inflation action, which has the potential,
\beq \label{eq:HiggsPotential}
V = \frac{\lambda}{4} \rpar{h^2 - v^2}^2 \approx \frac{\lambda}{4} h^4 \, ,
\eeq
where $v$ is the Higgs VEV, that can be neglected during Higgs inflation where $h \gg v$, such that a simple quartic potential is a good approximation. The prefactor $\lambda$ is fixed by experimentally measured values of $v$ and the mass of the Higgs boson. Higgs inflation also corresponds to the choice $F(h) = 1 + f h^2$ where $f$ is a constant parameter\footnote{While often called $\xi$ in the literature, we instead choose this notation to avoid confusion with the Gauss-Bonnet coupling function which is unfortunately also typically denoted $\xi(\phi)$ in the literature.}.

Note the presence of the constant $\omega$ multiplying the scalar field kinetic term. This is generally included in models containing the Gauss-Bonnet term as a constant which can be chosen to take the values $\pm 1$, the negative choice being necessary for certain forms of the Gauss-Bonnet coupling function to ensure conventional  inflationary behaviour \cite{Guo:2009uk}\cite{Guo:2010jr}. The standard model Higgs boson has a conventional kinetic term and thus $\omega$ is set equal to $1$ when we specialise to the Higgs case detailed above, but in the interest of maintaining greater generality we will work with unspecified values of $\omega$ as well as the three functions $V$, $F$, and $G$ in our derivations as much as possible. 

Usually, with non-minimally coupled models like standard Higgs inflation, one would usually proceed to now make a conformal transformation to the Einstein frame to reduce the problem to a more-studied, mathematically simpler problem with well-known solutions . However, the introduction of the Gauss-Bonnet term significantly complicates this approach. To see this, note that while under a conformal transformation of the form,
\beq \label{eq:conformaltrans}
g_{\mu \nu} \rightarrow e^{2\Gamma(h)} g_{\mu \nu}\, ,
\eeq
the Ricci scalar term in a standard Jordan frame theory (eq. (\ref{eq:gbhaction}) without the Gauss-Bonnet term) will look like
\beq
S \supset \int \bd^d x \sqrt{-g}\spar{\frac{1}{2} F e^{(d-2)\Gamma} R}  \, \eqtxt{(4D)} \, \int d^4 x \sqrt{-g} \spar{\frac{1}{2} F e^{2\Gamma} R} \, ,
\eeq
and it is simple to choose $2\Gamma = - \ln \rpar{F}$ to obtain the Einstein frame action. In contrast to this, under a conformal transformation the Gauss-Bonnet term becomes \cite{Carneiro:2004rt},
\begin{align}
E \rightarrow e^{-4 \Gamma}\, \Big\{ E &  - 8(d-3) R_{\mu \nu} \rpar{\Gamma^{;\mu}\Gamma^{;\nu} - \Gamma^{;\mu ;\nu}} - 2(d-3)R \left(2 \Box \Gamma + (d-4) \covd{\Gamma}^2 \right) \nonumber \\
 & + 4(d-2)(d-3) \left[ \cbbox{\Gamma}^2 + (d-3) \covd{\Gamma}^2 \cbbox{\Gamma} - \rpar{\Gamma_{;\mu ;\nu}\Gamma^{;\mu ;\nu} - 2 \Gamma_{;\mu;\nu}\Gamma^{;\mu}\Gamma^{;\nu}} \right] \nonumber \\
 & + (d-1)(d-2)(d-3)(d-4) \covd{\Gamma}^4 \Big\} \, ,
\end{align}
such that the full action including the Gauss-Bonnet term, following a conformal transformation, contains the term,
\beq
S \supset \int d^d x \sqrt{-g}\Big\{ F e^{(d-2)\Gamma} + 4 G e^{(d-4)\Gamma} \cbbox{\Gamma} \Big\} \frac{R}{2}  \, \eqtxt{(4D)} \, \int d^4 x \sqrt{-g}  \Big\{ F e^{-2\Gamma} + 4 G \cbbox{\Gamma} \Big\} \frac{R}{2} \, ,
\eeq
which complicates the usually simple choice of a function $\Gamma$ to implement the Einstein frame, as now the prefactor of $R$ includes $\cbbox{\Gamma}$ and the condition to set the coefficient of $R$ to a constant is dynamical. Further complications to the usual idea of the Einstein frame come from terms in the transformation such as,
\beq
S \supset \int d^d x \sqrt{-g} \spar{ 4 (d-3) e^{(d-4)\Gamma} G  R_{\mu \nu} \rpar{\Gamma^{;\mu}\Gamma^{;\nu} - \Gamma^{;\mu ;\nu}}} \, \eqtxt{(4D)} \,  \int d^4 x \sqrt{-g} \spar{4 G R_{\mu \nu} \rpar{\Gamma^{;\mu}\Gamma^{;\nu} - \Gamma^{;\mu ;\nu}}} \, .
\eeq
While all of these terms generated in the conformal transformation of the Gauss-Bonnet term are components of a general Hordenski theory, the complications involved in the analysis of this render the problem more tractable if we stay in the Jordan frame in which the model was formulated.

\section{Higgs Inflation in the Jordan Frame} \label{sec:jfhinflation}

To set the scene, we begin by retrieving the known results for the leading order slow-roll predictions of standard Higgs inflation, but without transforming the theory to the Einstein frame as is conventionally done. This is mainly to verify that our methods in the Jordan frame correctly reproduce the findings carried out in the Einstein frame, widely available in the literature, in anticipation of studying the full model including the Gauss-Bonnet term in the Jordan frame in the next section. Of course, in this we assume that choice of conformal frame does not alter the theory, as has been widely reaffirmed in the literature (see e.g. \cite{EFJF2,EFJF3,EFJF5}), at least until quantum corrections are considered \cite{Steinwachs:2013tr}.

We begin with the special case of the action in eq. (\ref{eq:gbhaction}) in which the Gauss-Bonnet coupling $G(h)$ is zero\footnote{More generally, we can study the case where $G(h)$ is a constant and obtain the same results, as the uncoupled Gauss-Bonnet term (and hence any constant multiple of it) does not contribute to the equations of motion in four dimensions. Only by choosing a non-constant $G(h)$ does this term actually modify the theory in a non-trivial way. } and $\omega = 1$ that is, 

\beq \label{eq:jfhaction}
S = \int \bd^d x \sqrt{-g} \left[ \frac{1}{2}F(h) R - \frac{1}{2} (\partial h)^2 - V(h) \right] \, ,
\eeq
which leads to the equations of motion 
\begin{align}
3 F H^2 & = \frac{1}{2} \dot{h}^2 + V(h) - 3 H \dot{F}\, , \label{eq:JFEE1}\\
2 F \dot{H} & = - \dot{h}^2 - \ddot{F} + H \dot{F} \, , \label{eq:JFEE2}\\ 
\ddot{h} + 3 H \dot{h} & + V_{,h} - 3(\dot{H} + 2 H^2) F_{,h} = 0 \label{eq:JFKG} \, .
\end{align}
While we have not chosen notation that explicitly reflects this, it is important to note that the variables we are referring to in these equations are Jordan frame variables. Due to the conformal factor relating the metrics of two frames, the scale factors and hence derived quantities such as the Hubble parameters $H$ and the e-fold count $N$ of the two frames differ \cite{EFJF1}. In particular, the e-fold count differs by a logarithmic correction $\Delta N \propto \ln\rpar{F(h)}$ \cite{kagan2008quantum,White:2013ufa}. Conventional Higgs inflation is typically formulated in the Einstein frame, and it is usually taken that 60 e-folds of Einstein frame expansion are required. The Jordan frame, is, however, the frame in which the couplings between the Higgs and the rest of the standard model are in their conventional form, so one may argue that it is the more correct frame to require 60 e-folds of inflation in. Nevertheless, even for a fairly large non-minimal coupling function $F$, the logarithmic difference in $N$ between the frames is going to be small. Indeed, we find at the level of a leading-order slow-roll analysis, neglecting this does not alter our results. We shall hence proceed to ignore this technicality and directly compare results in the two frames. 

Proceeding to calculate the number of remaining e-folds before the end of inflation, while neglecting terms proportional to e.g. $\ddot{h}$ to implement a slow-roll approximation, we obtain,
\beq \label{eq:efoldj}
N \approx \int_h^{h_f} \spar{ \frac{V}{2F}\rpar{\recfrac{2 V F_{,h} -  V_{,h} F}{2 F + 3 F_{,h}^2}} - \frac{F_{,h}}{F}} dh \approx \frac{3}{4}\rpar{1 + \frac{1}{6f}}\rpar{1 + f h^2} + \frac{1}{4} \ln\rpar{1 + f h^2} \, ,
\eeq
where in the second equality we have used the particular forms of $F$ and $V$ pertinent to Higgs inflation, as well as making the approximations $h \gg h_f$ and $h \gg v$ to keep things neat. If we then approximate this further by ignoring the logarithmic term (which is related to the difference between frames) and taking $f \gg 1$, we simplify our result to,
\beq \label{eq:JFefolds}
N \approx \frac{3}{4}\rpar{1 + f h^2} \, .
\eeq
This is equivalent (as discussed, up to the slight difference in meaning due to the conformal relation of the frames) to the Einstein frame result found ubiquitously in the literature \cite{Bezrukov:2007ep,Martin:2013tda}, as can be seen explicitly by considering the relationship between the Jordan frame field $h$ and the canonically normalised Einstein frame field $\chi$ \cite{Bezrukov:2013fka}, given by 
\beq
\rpar{\frac{\partial \chi}{\partial h}}^2 = \frac{ 1 + f \rpar{1+ \alpha^2 f} h^2}{\rpar{1 + f h^2}^2} \, .
\eeq

\subsection{Slow-roll in the Jordan frame}
While in standard (EF) inflation, it is useful to define a series of slow-roll parameters such as,
\beq \label{eq:standardsrp}
\epsilon_0 = \frac{-\dot{H}}{H^2} \, , \quad \epsilon_n = \frac{\dot{\epsilon}_{n-1}}{H \epsilon_{n-1}} \, ,
\eeq
which must satisfy $\eps{n} < 1$ to represent a period of sustained inflation, it is useful in the Jordan frame to define a second such series in terms of $F$. Noting that the equations of motion for this system give us an expression for $\epsilon_0$ of the form,
\beq \label{eq:JFepsilonform}
\epsilon_0 = \frac{\dot{h}^2}{2 H^2 F} + \frac{\ddot{F}}{2 H^2 F} -\frac{\dot{F}}{2 H F} \, ,
\eeq
we are motivated to define these Jordan frame slow-roll functions, which we will call $\zet{n}$ and similarly require their magnitude be small, such that,
\beq \label{eq:jfsrp}
\zeta_0 = \frac{\dot{F}}{H F} \, , \quad \zeta_n = \frac{\dot{\zeta}_{n-1}}{H \zeta_{n-1}} \, .
\eeq
This choice of slow-roll parameter is also motivated by noting that slow-roll can be thought of as the conditions $\ddot{h} \ll H \dot{h} \ll H^2 h$, which in turn imply for a smooth function $F(h)$ that $\ddot{F} \ll H \dot{F} \ll H^2 F$ \cite{EFJF4}. We can then expand quantities in terms of a combination of the conventional $\eps{n}$ parameters and these new $\zet{n}$ functions as in standard Einstein frame models to construct a slow-roll analysis of a model in the Jordan frame.  It is helpful to expand the kinetic and potential terms for the $h$ field in terms of slow-roll parameters, and find that,
\begin{align}
\dot{h}^2 &=    H^2 F \bigg[ 2 \eps{0} + \zet{0} - \zet{0}\rpar{\zet{1} + \zet{0} - \eps{0}} \bigg] \label{eq:SRkinetic} \, , \\
V &=  H^2 F \spar{3 - \eps{0} + \frac{5}{2}\zet{0} + \frac{1}{2}\zet{0}\rpar{\zet{1} + \zet{0} - \eps{0}}} \label{eq:SRpotential} \, .
\end{align}
Some useful relations for performing slow-roll expansions in this formalism are given in appendix \ref{app:GBHSRA}.
Following a full numerical integration of these equations of motion to confirm reasonable inflationary solutions exist and conform roughly to our expectations (such as approximate number of e-folds) to verify our methods so far, we proceed to analyse the spectrum of perturbations at first order in slow-roll in order to compare them to the Einstein frame formulation of the theory.

\subsection{Scalar Perturbations}
The linear perturbation equations for this system can still be written in Sasaki-Mukhanov form despite the non-minimal coupling function (we will later see that adding the Gauss-Bonnet term also does not prevent us from transforming the perturbation equations into such a form) \cite{StringGaussBonnet1,Hwang:1999gf}. That is, working in terms of the variable $\upsilon_s = z_s \mathcal{R}$ where $\mathcal{R}$ is the curvature perturbation, we have the equation of motion for a given mode $k$,
\beq \label{eq:HiggsJFSM}
\upsilon_s'' + \rpar{k^2 - \frac{z_s''}{z_s}} \upsilon_s = 0 \, ,
\eeq
where $z_s = a \sqrt{Q_s}$, the prime represents differentiation with respect to conformal time coordinate $\eta$, satisfying $\bd t = a \bd \eta$, and 
\beq \label{eq:HiggsJFQs}
Q_s = \frac{\dot{h}^2 + \frac{3}{2} \frac{\dot{F}^2}{F}}{\rpar{H + \frac{1}{2}\frac{\dot{F}}{F}}^2} \, .
\eeq
The sound speed of the perturbations meanwhile remains as $c_s = 1$ in the Jordan frame. Exact solutions of this are linear combinations of Hankel functions such that,
\beq
\upsilon_s = \sqrt{-\eta} \rpar{C_1 H^{(1)}_{\Omega_s} (- k \eta) + C_2 H^{(2)}_{\Omega_s} (- k \eta)} \, ,
\eeq
where $\Omega_s^2 = 1/4 + \eta^2 z_s''/z_s$, which can be systematically, albeit tediously, expanded in slow-roll parameters as necessary.

Quantisation of the perturbations requires the solution to asymptotically tend towards the Bunch-Davies initial condition at early times ($\eta \rightarrow - \infty$), we find the integration constants must be $C_1 = \sqrt{\pi/4} \exp (i \pi \rpar{\Omega + 1/2}/2)$ and $C_2  = 0$. The solution at late times ($\eta \rightarrow 0$) is then of the form,
\beq
\upsilon_s = \spar{2^{\Omega_s - 3/2} \frac{\Gamma(\Omega_s)}{\Gamma(\frac{3}{2})}} e^{\rpar{ \frac{i \pi}{2} \rpar{\Omega_s + \frac{1}{2}}}} \sqrt{\frac{\eta}{2}} \rpar{-k \eta}^{-\Omega_s} \, ,
\eeq
and has a corresponding power spectrum,
\begin{align}
\mathcal{P}_\mathcal{R}  = & \ \frac{k^3}{2 \pi^2} \left|\frac{\upsilon_s}{z_s}\right|^2 = \frac{k^3}{2 \pi^2 a^2 } \frac{\left|\upsilon_s\right|^2}{\left|Q_s\right|} \, , \nonumber \\
= & \ \spar{2^{\Omega_s - 3/2} \frac{\Gamma(\Omega_s)}{\Gamma(\frac{3}{2})}}^2 \frac{H^2}{4 \pi^2 \rpar{a H \eta}^2 \left|Q_s\right|} \rpar{k \eta}^{3-2\Omega_s} \label{eq:JFgeneralscalarspectrum}\, .
\end{align}
Following the usual procedure of evaluating this spectrum at the time when $k = a H$ (when the mode $k$ is crossing the horizon) and approximating the spectrum to leading order in slow-roll, we find
\beq 
\mathcal{P}_\mathcal{R} \approx \frac{H^2}{4 \pi^2 \left|Q_s\right|} \, .
\eeq
Using the dual slow-roll expansion in $\eps{0}$ and $\zet{0}$ discussed above we can re-express $Q_s$ in terms of slow-roll quantities as
\beq
Q_s =  \frac{F A}{\rpar{1 + \frac{1}{2} \zet{0}}^2} \, ,
\eeq
where
\beq
A = 2 \eps{0} + \zet{0} - \zet{0}\rpar{\zet{1} - \eps{0}} + \frac{1}{2} \zet{0}^2 \, .
\eeq
In terms of slow-roll parameters, we then have at leading order,
\beq \label{eq:JFslowrollspectrum}
\mathcal{P}_\mathcal{R} \approx \frac{H^2}{4 \pi^2 F \left|2 \eps{0} + \zet{0}\right|} \, .
\eeq
This expression looks qualitatively similar to the standard Einstein frame result, indeed, setting $F = 1$ (and hence $\zet{0} \propto \dot{F} = 0$ we recover the usual result. Now, specifying to the appropriate potential and non-minimal coupling function for Higgs inflation, we can evaluate this spectrum for a given parameter choice, and find the scalar amplitude can be written in the form,
\beq \label{eq:JFamplitude}
A_s = \frac{\lambda h^4}{128 \pi^2} = \frac{\lambda N^2 }{72 \pi^2 f^2} \, ,
\eeq
where in the second equality we have used eqn. (\ref{eq:JFefolds}) and $f \gg 1$ to rewrite the amplitude as a function of number of e-folds remaining before inflation ends. This result is in agreement with the Einstein frame version of the calculation \cite{Bezrukov:2013fka}, and hence with $N = 60$, also requires $f \approx 50000 \sqrt{\lambda}$ for normalisation to the CMB data. 

We can similarly evaluate the spectral index in a slow-roll expansion, to find at leading order that,
\beq
n_s - 1 = - 2 \eps{0} - \frac{2 \eps{0}\rpar{\zet{0}+\eps{1}} + \zet{0}\rpar{\zet{0}+\zet{1}}}{2\eps{0}+\zet{0}} \, . \\
\eeq
This, evaluated using the relevant functions for Higgs inflation, takes the form,
\beq
n_s - 1 \approx -\frac{8N + 7}{4 N^2} \approx  \frac{-2}{N} \, ,
\eeq
which, again, agrees with the Einstein frame result at first order \cite{Bezrukov:2013fka} and for $N = 60$ gives $n_s \approx 0.967$. The close equivalence of these spectral properties has also been observed numerically, using a full integration of the background equations of motion in each frame, with the small observed differences (typically at the 3rd or 4th significant figure) being attributed to approximating $N$ to be the same in both frames.
\subsection{Tensor Perturbations}
Similarly analysing the spectrum of tensor perturbations in the Jordan frame, we once again write the perturbation equation for a single mode in Sasaki-Mukhanov form,
\beq \label{eq:HiggsJFSMt}
\upsilon_t'' + \rpar{k^2 - \frac{z_t''}{z_t}} \upsilon_t = 0 \, ,
\eeq
where $\upsilon_t$ is related to the tensor mode by a factor of $z_t = a Q_t$, where in this case $Q_t$ is simply $F$. The full gravitational wave spectrum is then, in analogy with the solution given for scalar modes in the previous section,
\begin{align}
\mathcal{P}_\mathcal{T}  = & \ \frac{2 k^3}{2 \pi^2} \left|\frac{2 \upsilon_t}{z_t}\right|^2 = \frac{8 k^3}{2 \pi^2 a^2 } \frac{\left|\upsilon_t\right|^2}{\left|Q_t\right|} \, , \nonumber \\
= & \ \spar{2^{\Omega_t} \frac{\Gamma(\Omega_t)}{\Gamma(\frac{3}{2})}}^2 \frac{H^2}{4 \pi^2 c_t^3 \rpar{a H \eta}^2 \left|Q_t\right|} \rpar{c_t k \eta}^{3-2\Omega_t} \, ,
\end{align}
where $\Omega_t^2 = 1/4 + \eta^2 z_t''/z_t$. This can then be approximated (again at the point when $k = a H$) as,
\beq 
\mathcal{P}_\mathcal{T} \approx \frac{2 H^2}{\pi^2 \left|Q_t\right|} \, ,
\eeq
which in combination with the scalar amplitude calculated in the previous section lets us obtain the tensor-to-scalar-ratio,
\beq \label{eq:JFttsr}
r = \frac{\mathcal{P}_\mathcal{T}}{\mathcal{P}_\mathcal{R}} = 8 \left|\frac{Q_s}{Q_t}\right| \approx \frac{12}{N^2} \, .
\eeq
Once again this is consistent with the Einstein frame version of the theory \cite{Bezrukov:2013fka}, and has also been verified numerically. Furthermore we find that $r = - 8 n_t$, replicating the consistency relation. Given these results as well as those of the scalar spectrum in the previous section, we are now confident that working in the Jordan frame with these particular methods is able to  correctly rediscover the results of usual Higgs inflation. While the calculations for this particular example are only complicated by the choice of the Jordan frame, we will now go on to apply these tested methods to the full action containing the Gauss-Bonnet term which does not behave simply under conformal frame transformations.

\section{Gauss-Bonnet-Higgs Inflation} \label{sec:GBHinflation}

We now proceed to study the full action for the theory, as shown in eq. (\ref{eq:gbhaction}), using the Jordan frame techniques developed and tested in the previous section. Firstly, the background equations of motion for Gauss-Bonnet-Higgs (GBH) inflation are found by variation of the action to be
\begin{align}
3 H^2 (F - 4 H \dot{G})& = \frac{1}{2} \omega \dot{h}^2 + V(h) - 3 H \dot{F}\, , \label{eq:GBHEE1} \\
2(F - 4 H \dot{G}) \dot{H} & = - \omega \dot{h}^2 - \ddot{F} + H \dot{F} + 4 H^2 (\ddot{G}-H\dot{G}) \, \label{eq:GBHEE2} ,\\ 
\omega (\ddot{h} + 3 H \dot{h}) & + V_{,h} - 3(\dot{H} + 2 H^2) F_{,h} + 12 H^2 G_{,h} (\dot{H} + H^2) = 0 \label{eq:GBHKG} \, .
\end{align}
We can use this to construct an integral to approximate the number of e-folds of inflation produced, and find that,
\beq \label{eq:efoldgen}
N = \int_h^{h_f} \spar{ \frac{V}{2F^2}\rpar{\recfrac{6 V F_{,h} F - 3 V_{,h} F^2 - 4 G_{,h} V^2}{6 \omega F^3 + 9 F_{,h}^2 F^2 - 24 G_{,h} F_{,h} V F + 16 G_{,h}^2 V^2}} - \frac{F_{,h}}{F} + \frac{4}{3} \frac{G_{,h} V}{F^2}} dh \, .
\eeq
Furthermore, we can define slow-roll parameters for this model. As in the case of standard Jordan frame inflation, we have the $\epsilon_{n}$ (eq. (\ref{eq:standardsrp})) and $\zet{n}$ (eq. (\ref{eq:jfsrp})) functions, but the presence of the Gauss-Bonnet coupling further inspires the definition of another set of functions we will denote by $\del{n}$, which have the form \cite{Guo:2010jr}
\beq
\delta_0 = 4 \dot{G} H \, , \quad \delta_n = \frac{\dot{\delta}_{n-1}}{H \delta_{n-1}} \, .
\eeq
This form of the Gauss-Bonnet slow-roll functions has been used in previous studies of inflation in the presence of the Gauss-Bonnet term coupled to the Einstein frame field. However, we find that the combination of both a Gauss-Bonnet coupling $G(h)$ and a non-minimal coupling function $F(h)$, the first standard slow-roll parameter $\eps{0}$ can be expressed using eqs. (\ref{eq:GBHEE1}) -- (\ref{eq:GBHKG}) as,
\beq \label{eq:GBHepsilonform}
\epsilon_0 = \frac{\frac{\omega \dot{h}^2}{H^2 F} + \frac{\ddot{F}}{H^2 F} -\frac{\dot{F}}{H F} - \frac{4 \rpar{\ddot{G}-H\dot{G}}}{F}}{2 \rpar{1 - \frac{4 H \dot{G}}{F}}} \, .
\eeq
In order for $\eps{0}$ to be small, the denominator of this expression must not approach zero, which in turn implies that we require $4 H \dot{G} / F \ll 1$ to sustain inflation. This, rather than the condition in the Einstein frame, that is,  $\del{0} = 4 H \dot{G} \ll 1$, is hence more suited as a basis for defining a slow-roll parameter for this model. Therefore instead of the $\del{n}$ functions, we define and will use modified Gauss-Bonnet flow functions that satisfy,
\beq
\Delta_0 = \frac{\delta_0}{F} = \frac{4 H \dot{G}}{F}  \, , \quad \Delta_n = \frac{\dot{\Delta}_{n-1}}{H \Delta_{n-1}} \, .
\eeq
Useful relations for manipulating these different types of slow-roll parameters have been compiled in appendix \ref{app:GBHSRA}.  As well as more accurately representing the required condition to achieve slow-roll inflation in the Jordan frame, the use of the $\Del{n}$ parameters also simplifies expressions written in terms of slow-roll parameters. For example, expanding the kinetic and potential terms in slow-roll, we find
\begin{align}
\omega \dot{h}^2 &=   H^2 F \bigg[2 \eps{0} + \zet{0} - \Del{0} - \zet{0}\rpar{\zet{1} + \zet{0} - \eps{0}} + \Del{0}\rpar{\Del{1} + \zet{0} - \eps{0}}\bigg] \label{eq:GBHSRkinetic} \, , \\
V(h) &=  H^2 F \spar{3 - \eps{0} + \frac{5}{2}\zet{0} + \frac{1}{2} \Del{0} + \frac{1}{2}\zet{0}\rpar{\zet{1} + \zet{0} - \eps{0}} - \frac{1}{2} \Del{0}\rpar{\Del{1} + \zet{0} - \eps{0}}} \label{eq:GBHSRpotential} \, .
\end{align}
Had we expanded the above in terms of $\del{n}$ parameters instead, not all terms in the expansion would be the same order in $F$. In this sense, the $\Del{n}$ parameters are again the more appropriate choice as they appear in expressions on the same footing as the other kinds of slow-roll functions. 

\subsection{Scalar Perturbations}
Once again we are able to cast the scalar perturbation equation in the simple Sasaki-Mukhanov form \cite{Hwang:1999gf},
\beq \label{eq:GBHSM}
\upsilon_s'' + \rpar{c_s^2 k^2 - \frac{z_s''}{z_s}} \upsilon_s = 0 \, ,
\eeq
albeit now with a non-trivial sound-speed given by,
\beq \label{eq:GBHcs}
c_s^2 = 1 + 4 \dot{G} \frac{\frac{1}{2} \rpar{\frac{\dot{F} - 4 H^2 \dot{G}}{F - 4 H \dot{G}}}^2 \rpar{\frac{\ddot{G}}{G} - H - 4 \dot{H} \frac{F - 4 H \dot{G}}{\dot{F} - 4 H^2 \dot{G}} }}{\omega \dot{h}^2 + \frac{3}{2} \frac{\rpar{\dot{F} - 4 H^2 \dot{G}}^2}{F - 4 H \dot{G}} } \, ,
\eeq
and as before we have $z_s = a Q_s$ with $Q_s$ given by
\beq \label{eq:GBHQ}
Q_s = \frac{\omega \dot{h}^2 + \frac{3}{2} \frac{\rpar{\dot{F} - 4 H^2 \dot{G}}^2}{F - 4 H \dot{G}}}{\rpar{H + \frac{1}{2}\frac{\dot{F} - 4 H^2 \dot{G}}{F - 4 H \dot{G}}}^2} \, .
\eeq
Expanded in slow roll parameters, this can be written as
\beq
Q_s =  \frac{F A}{\rpar{1 + \frac{1}{2} x}^2} \, ,
\eeq
where
\beq
A = 2 \eps{0} + \zet{0} - \Del{0} - \zet{0}\rpar{\zet{1} + \zet{0} - \eps{0}} + \Del{0}\rpar{\Del{1} + \zet{0} - \eps{0}} + \frac{3}{2} \rpar{\zet{0} - \Del{0}} x \, ,
\eeq
and
\beq
x = \frac{\zet{0} - \Del{0}}{1 - \Del{0}} \, .
\eeq
Meanwhile $c_s$ has a slow-roll expansion of,
\beq
c_s^2 = 1 + \frac{x B}{2 A} \, ,
\eeq
with $B$ defined as,
\beq
B = \Del{0} \rpar{4 \eps{0} - x + x \spar{\Del{1} + \zet{0} + \eps{0} }} \, .
\eeq
We note from these expressions that $Q_s = \mathcal{O}(\epsilon)$, $c_s^2 = 1 + \mathcal{O}(\epsilon^2)$. That is, even though we have a non-trivial expression for the sound speed in principle, its deviation from unity is second order in slow-roll, and hence in practice only plays a small role in inflation.  

We can proceed to compute the scalar spectrum in the usual way, using the Hankel-function solution of eq. (\ref{eq:GBHSM}) discussed in the previous section, to obtain the expression,
\begin{align}
\mathcal{P}_\mathcal{R}  = & \ \frac{k^3}{2 \pi^2} \left|\frac{c_s \upsilon_s}{z_s}\right|^2 = \frac{k^3}{2 \pi^2 a^2 } \frac{\left| c_s \upsilon_s\right|^2}{\left|Q_s\right|} \, , \nonumber \\
= & \ \spar{2^{\Omega_s - 3/2} \frac{\Gamma(\Omega_s)}{\Gamma(\frac{3}{2})}}^2 \frac{H^2}{4 \pi^2 c_s \rpar{a H \eta}^2 \left|Q_s\right|} \rpar{c_s k \eta}^{3-2\Omega_s} \label{eq:GBHscalaranalytics} \, .
\end{align}
Evaluating this in terms of slow-roll parameters at the moment when $c_s k = a H$ and retaining only the leading order terms we find the results
\beq \label{eq:GBHscalarspec}
A_s \approx \frac{H^2}{4 \pi^2 \left|Q_s\right|} \approx \frac{H^2}{4 \pi^2 F \left|2 \eps{0} + \zet{0} - \Del{0}\right|} \, ,
\eeq
for the scalar amplitude and,
\beq
n_s = 1 - 2 \eps{0} - \frac{2 \eps{0}\rpar{\zet{0}+\eps{1}} + \zet{0}\rpar{\zet{0}+\zet{1}} - \Del{0}\rpar{\zet{0}-\Del{1}} }{2\eps{0}+\zet{0}-\Del{0}} \, ,
\eeq
for the corresponding spectral index. 

\subsection{Tensor Perturbations}
The equation of motion for tensor perturbations in this model can also be expressed in the form,
\beq \label{eq:GBHSMt}
\upsilon_t'' + \rpar{c_t^2 k^2 - \frac{z_t''}{z_t}} \upsilon_t = 0 \, ,
\eeq
with sound speed
\beq
c_t^2 = \frac{F - 4 \ddot{G}}{F - 4 H \dot{G}} \, ,
\eeq
and $z_t = a \sqrt{Q_t}$, with $Q_t$ found to be,
\beq
Q_t = F - 4 H \dot{G} \, .
\eeq
These expressions can then be written in terms of the slow-roll parameters as,
\beq
Q_t = F \rpar{1 - \Del{0}} \, ,
\eeq
and
\beq
c_t^2 = \frac{1 - \Del{0}\rpar{\Del{1}+\zet{0}+\eps{0}}}{1 - \Del{0}} \, .
\eeq
Note that for the tensor spectrum, the squared sound speed now deviates from unity at first order in slow-roll. Following the usual method once again, we can write down the tensor power spectrum,
\begin{align}
\mathcal{P}_\mathcal{T}  = & \ \frac{2 k^3}{2 \pi^2} \left|\frac{2 c_t \upsilon_t}{z_t}\right|^2 = \frac{8 k^3}{2 \pi^2 a^2 } \frac{\left| c_t \upsilon_t\right|^2}{\left|Q_t\right|} \, , \nonumber \\
= & \ \spar{2^{\Omega_t} \frac{\Gamma(\Omega_t)}{\Gamma(\frac{3}{2})}}^2 \frac{H^2}{4 \pi^2 c_t \rpar{a H \eta}^2 \left|Q_t\right|} \rpar{c_t k \eta}^{3-2\Omega_t} \, , \label{eq:GBHtensoranalytics}
\end{align}
and then evaluate it at the time when $c_t k = a H$, neglecting higher order terms in the slow-roll expansion, to find that at leading order the tensor-to-scalar ratio is,
\beq \label{eq:GBHttsr}
r = \frac{\mathcal{P}_\mathcal{T}}{\mathcal{P}_\mathcal{R}} = 8 \left|\frac{Q_s}{Q_t}\right| = 8 \frac{2 \eps{0} + \zet{0} - \Del{0}}{1 - \Del{0}} \, .
\eeq
Note that in doing this we have also ignored the fact that as $c_s \neq c_t$, scalar and tensor modes do not exit the horizon simultaneously. We also find the leading order tensor spectral index in this model to be independent of the Gauss-Bonnet correction, taking the form,
\beq
n_t  = - 2 \eps{0} - \zet{0} \, ,
\eeq
which means that $r \neq - 8 n_t$, breaking the leading order consistency relation of standard inflation, as is also the case for Gauss-Bonnet-coupled inflation in the Einstein frame.

\section{Results}

Having now derived some generic results for scalar-tensor theories in the Jordan frame with a Gauss-Bonnet coupling, we are now in a position to specialise to particular choices of coupling function to investigate this model in detail. As discussed we are particularly interested in Higgs inflation so we use $F(h) = 1 + f h^2$ with $f$ chosen to normalise the amplitude of the scalar spectrum and the Higgs potential of eq. (\ref{eq:HiggsPotential}). For simplicity, and to gain an idea of what kind of generic effects the Gauss-Bonnet term has during inflation, we typically parametrise it as a power law $G = G_0 h^g$ for this work, particularly focusing on the case where $g = -4$, unless otherwise stated, as this simplifies many of the resulting equations.

Our results are obtained via numerical integration of the system of equations specified by eqs. (\ref{eq:GBHEE1})--(\ref{eq:GBHKG}). From this, all quantities defined at the background level are extracted and used to evaluate the analytical expressions for the spectra derived in section \ref{sec:GBHinflation} (see eq. (\ref{eq:GBHscalaranalytics}) for scalars and eq. (\ref{eq:GBHtensoranalytics}) for tensors).

\subsection{Jordan Frame Higgs Inflation with a Gauss-Bonnet modification} \label{sec:GBHstable}

Numerically we find the surprising result, that, without choosing excessively large values for the parameters $G_0$ and $g$ in the Gauss-Bonnet coupling, no significant modifications to the standard Higgs inflation spectra occur. While in general it is possible to create inflationary models where the coupling to the Gauss-Bonnet term has a prominent role in the dynamics and hence strongly determines the properties of the primordial spectra, the properties of Higgs inflation naturally suppress this effect. To see this we first note that during the observable window of Higgs inflation when slow-roll is applicable we have $H \approx \mathcal{O}(10^{-6})$ and that in the equations of motion, terms involving the Gauss-Bonnet coupling typically appear alongside higher powers of $H$ than the rest of the terms. For example, in eq. (\ref{eq:GBHKG}) the Gauss-Bonnet contribution is suppressed by a factor of $H^4$ relative to the potential term. This, combined with the smallness of $H$, stabilises Higgs inflation under Gauss-Bonnet-like corrections for all but unnaturally large choices of coupling parameters. 

This is also apparent from the definition of the different kinds of slow-roll parameter (see previous sections or appendix \ref{app:GBHSRA}); the first Gauss-Bonnet slow-roll function $\Del{0}$ is proportional to $H$, which is small, and inversely proportional to $F$, which is always greater than 1 for positive $f$. Once again we see that unless $\dot{G}$ is especially large (at least $\mathcal{O}(H^{-1})$ or so)  we will have $\Del{0} \approx 0$ and the Gauss-Bonnet corrections to spectra will be negligible.

A further consideration is that especially for positive power-law couplings, as Higgs inflation has $h \ll 1$, large powers of $h$ in the coupling function will further suppress the magnitude of the $\Del{n}$ parameters. On a related note, while this in turn means that inverse-power-law type couplings to the Gauss-Bonnet term (which will be considered in further detail in section \ref{sec:GBend}) could feasibly produce a large enough value of $G_h$ as $h \rightarrow 0$ during inflation, this effect is only significant at the end of inflation, not in the much earlier observable window when the measured primordial spectra are determined, for reasonable choices of $g$. 

\subsection{Effects of Gauss-Bonnet on the end of inflation} \label{sec:GBend}

\begin{figure}[t]
    \centering
    \includegraphics[width=\textwidth]{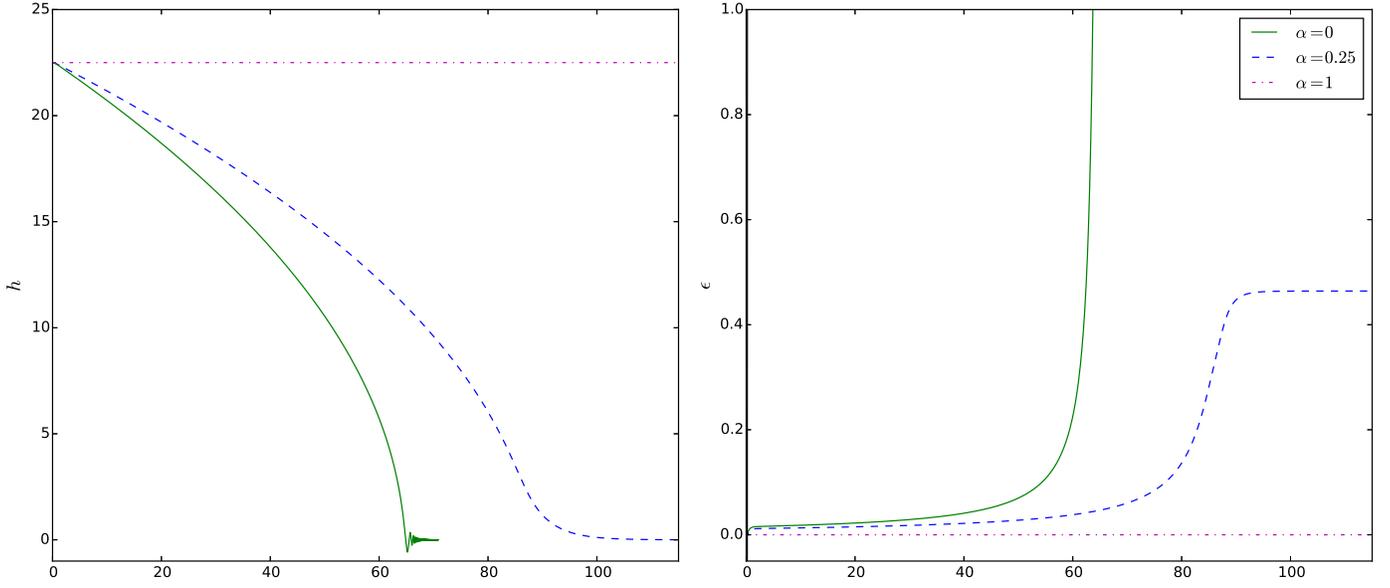}
    \caption{Numerical evolutions of $h$ (left) and $\eps{0}$ (right) for Gauss-Bonnet inflation with $F = 1$ ($f = 0$), $h_0 = 22.5$, and the Higgs potential and inverse power-law ($g = -4$) coupling to the Gauss-Bonnet term. The solid-green line corresponds to $\alpha = 0$ (standard Higgs inflation) and is shown for comparison with the other trajectories with non-trivial Gauss-Bonnet effects. Of these, the blue-dashed line has $\alpha = 0.25$ and shows late-time constant $\eps{0}$, while the purple-dotted line has $\alpha = 1$ which causes the scalar field to behave as a cosmological constant.}
    \label{fig:GB_latetime}
\end{figure}

\begin{figure}[t]
    \centering
    \includegraphics[width=\textwidth]{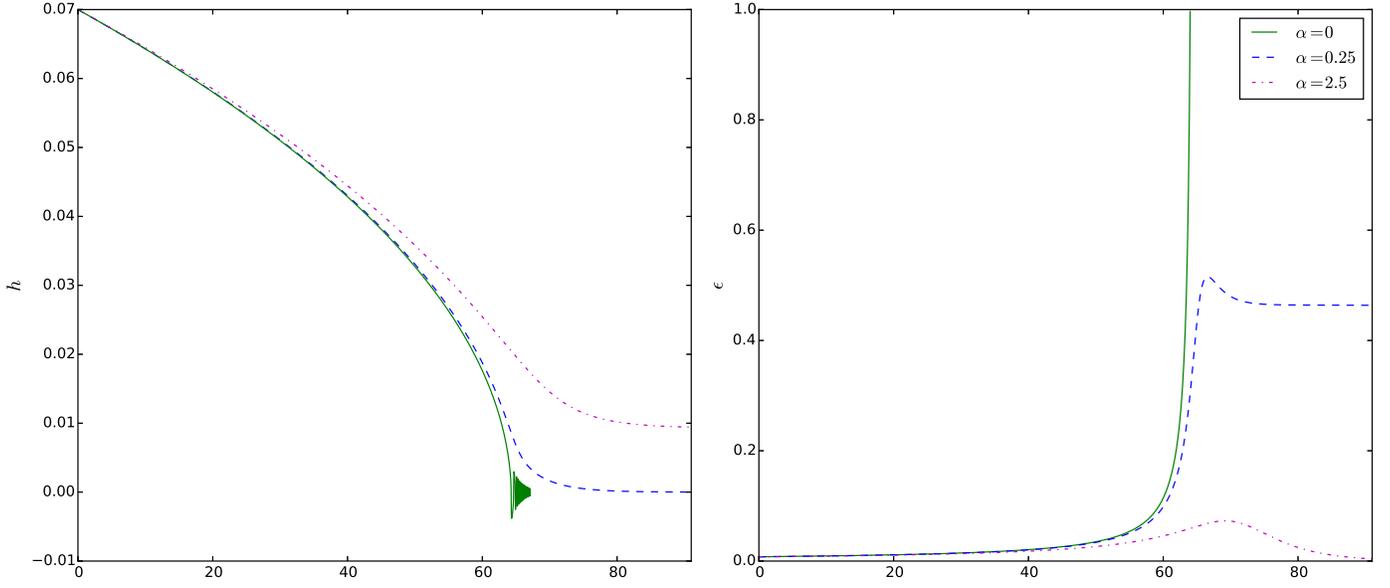}
    \caption{Numerical evolutions of $h$ (left) and $\eps{0}$ (right) with the choice of functions and parameters corresponding to standard Higgs inflation ($h_0 = 0.07$, $f = 17367.233$) and inverse power-law ($g = -4$) coupling to the Gauss-Bonnet term.. The solid-green line corresponds to $\alpha = 0$ (standard Higgs inflation) and is shown for comparison with the other trajectories with non-trivial Gauss-Bonnet effects. Of these, the blue-dashed line has $\alpha = 0.25$ and shows late-time constant $\eps{0}$, while the purple-dotted line has $\alpha = 2.5$ which causes the scalar field to behave as a cosmological constant at late times. Compared to trajectories shown in Fig. \ref{fig:GB_latetime} with $f = 0$, the primary effect of a non-zero $f$ is to affect the transition between the early-time behaviour, which resembles normal inflation, and the late time behaviour with constant $\eps{0}$. In particular, the transition between these two periods has a local feature in the evolution of $\eps{0}$. For the cosmological constant trajectory, this feature also manifests to temporarily bring about a small deviation from the otherwise de-Sitter-like expansion. }
    \label{fig:GB_latetime_F}
\end{figure}

With inverse power-law couplings, $G_h$ becomes very large towards the end of inflation as $h \rightarrow 0$ such that even in the presence of the aforementioned large suppressions proportional to powers of the Hubble parameter, interesting effects may occur. Numerical investigations further reveal that this manifests as the absence of late-time oscillations of the inflaton when $G$ becomes large enough, typically with an inverse-power-law coupling so that as $h \rightarrow 0$, $G \rightarrow \infty$. This occurs both in the absence of the non-minimal coupling function $F$ (see Fig. \ref{fig:GB_latetime}) and in its presence (see Fig. \ref{fig:GB_latetime_F}) such as is the case for Higgs inflation.  

In lieu of the usual oscillations, the late-time behaviour of the system is to approach a constant value of $\eps{0}$ dependent on the magnitude of the Gauss-Bonnet coupling (blue-dashed trajectories of Figs. \ref{fig:GB_latetime} and \ref{fig:GB_latetime_F}), in contrast to normal inflation in which the late-time behaviour corresponds to a rapid increase in $\eps{0}$ signifying the end of inflation (shown as green-solid trajectories in Figs. \ref{fig:GB_latetime} and \ref{fig:GB_latetime_F}). Furthermore, if the constant value of $\eps{0}$ at late times is $0$, then the inflaton/Higgs--field correspondingly approaches a constant, non-zero value, behaving like a cosmological constant (purple-dotted trajectories of Figs.  \ref{fig:GB_latetime} and \ref{fig:GB_latetime_F}). We leave to future work the study of the implications of this for post-inflationary cosmology, such as the implications for (p)reheating and whether this finding can realistically be used to explain the observed late-time accelerating expansion of the universe. For now, we proceed to study the existence and properties of these solutions. For this purpose it is convenient to define the parameter $\alpha$ as,
\beq
\alpha = \frac{4 V_0 G_0}{3} = \frac{\lambda G_0}{3} \, ,
\eeq
for power-law functions $V = V_0 h^n$ and $G = G_0 h^g$ and where, in the second equality, we have specialised to the Higgs potential with $V_0 = \lambda/4$. The late-time behaviour of the system is determined by the value of $\alpha$. In particular, let us rewrite eq. (\ref{eq:GBHKG}) with $\dot{h} = \ddot{h} = \dot{H} = 0$ and look for the aforementioned cosmological-constant solutions where $h = h_\Lambda$ is a constant. We find,

\beq
V_{,h}(h_\Lambda) F(h_\Lambda)^2 - 2 V(h_\Lambda) F(h_\Lambda) F_{,h}(h_\Lambda) + \frac{4}{3} V(h_\Lambda)^2 G_{,h}(h_\Lambda) = 0 \, ,
\eeq
in which we have used eq. (\ref{eq:GBHEE1}) to rewrite $H^2$ as $V/3F$ when $\dot{h}$, and hence $\dot{F}$ and $\dot{G}$ are zero. Specifying the forms of the three functions $V$, $F$, and $G$ we are concentrating on, this expression becomes

\beq \label{eq:constanthbound}
1+  f h_\Lambda^2 + \frac{1}{12} \lambda g G_0 h_\Lambda^{g+4} = 0 \, .
\eeq
First we note from this that for the case when $G_0 = 0$ or $g = 0$ (when the Gauss-Bonnet coupling has no effect) this reduces to $1 + f h^2 = 0$ which has no real solutions for positive $f$, as is the case with Higgs inflation. That is, without the Gauss-Bonnet term, we do not find a constant-$h$ solution. Secondly, studying the case when $f = 0$, we can rewrite the condition in eq. (\ref{eq:constanthbound}) as
\beq
1 + \frac{g}{4} \alpha h_{\Lambda}^{g+4} = 0 \, .
\eeq

In the simplest case, when $g = -4$, this becomes $1 - \alpha = 0$, that is, constant-$h$ solutions only exist for $\alpha = 1$. This solution is realised numerically and shown in Fig.  \ref{fig:GB_latetime} (dotted purple line). 

The full constraint in eq. (\ref{eq:constanthbound}) when $f \neq 0$ cannot generally be solved analytically, but again specialising to the case where $g = -4$ we can show that,
\beq \label{eq:hlambdaprediction}
h_\Lambda= \sqrt{\frac{\alpha - 1}{f}} \, ,
\eeq
from which we see that real values of $h_\Lambda$ obey the constraint equation when $\alpha \geq 1$. An example of this is shown in Fig. \ref{fig:GB_latetime_F} where $\alpha = 2.5$ and $f = 17367.233$, and hence the late-time constant value approached for this trajectory is given by eq. (\ref{eq:hlambdaprediction}) as $h_\Lambda = 0.0092935217$, in agreement with the numerical result presented. We note that in contrast to the pure Gauss-Bonnet case where $f = 0$ and $\alpha$ must be exactly $1$ to lead to cosmological constant solutions (and $\alpha > 1$ corresponds to non-inflationary trajectories with growing $h$), the additional presence of a simple but non-trivial coupling to the Ricci scalar here has increased the parameter space of cosmological constant solutions to all $\alpha \geq 1$. 

For $0 < \alpha < 1$, rather than cosmological constant-like behaviour ($\eps{0} \rightarrow 0$ at late times), we instead observe $\eps{0}$ approach a non-zero constant. In such cases, shown in the blue-dashed trajectories of Figs. \ref{fig:GB_latetime} and \ref{fig:GB_latetime_F}, we note that the value of $\eps{0}$ approached is essentially independent of $f$ and initial conditions, depending only on $\alpha$. This has been numerically verified for several cases, one example of which is shown in the $f=0$ example in Fig. \ref{fig:GB_latetime} and the large $f$ example in Fig. \ref{fig:GB_latetime_F}, which both have $\alpha = 0.25$ and approach $\eps{0} \approx 0.461$. During a period when $\eps{0} = \eps{c}$ is approximately constant, we can use the definition of $\eps{0}$ to write $\dot{H} = - \eps{c} H^2$, or alternatively, $H' = - \eps{c} H$ where $H'$ indicates the derivative with respect to e-fold number $N$. Solutions of this differential equation will in turn approximate the behaviour of $H$ during this period, for which we find 
\beq
H = H\!\rpar{N_0} \exp \rpar{ - \eps{c} \rpar{N - N_0}} \, ,
\eeq
where $N_0$ is the value of $N$ when $\eps{0}$ began to be (approximately) constant. Given the behaviour of $H$, we can then infer the evolution of $h$ through, say, eq. (\ref{eq:GBHSRpotential}), in the form,

\beq
H^2 = \frac{V}{\rpar{3 + \mathcal{O}(\eps{0})}F} \approx \frac{V}{3F} \approx \frac{\lambda h^4}{12 \rpar{1 + f h^2}} \, ,
\eeq
which is valid when the constant value of $\eps{0}$ is considerably less than $3$. From this, the evolution of $h$ is given by, 

\beq
    h \approx 
    \begin{cases}
      \sqrt{\frac{12 f}{\lambda}} H = \sqrt{\frac{12 f}{\lambda}} H\!\rpar{N_0} \exp \rpar{ -\eps{c} \rpar{N - N_0}}  ,& \quad f h^2 \gg 1 \, ,  \\
      \rpar{\frac{12}{\lambda}}^{1/4} \sqrt{H} = \rpar{\frac{12 H\!\rpar{N_0}^2}{\lambda}}^{1/4} \exp \rpar{ - \eps{c} \rpar{N - N_0}/ 2},&  \quad f h^2 \ll 1  \, .
    \end{cases}
\eeq

Typically at late times (or always, when $f = 0$) it is the latter case ($fh^2 \ll 1$) that will be relevant. In either case $h$ exponentially declines towards $0$, which can be seen qualitatively in the blue-dashed trajectories of Figs. (\ref{fig:GB_latetime}) and (\ref{fig:GB_latetime_F}). 

We finally emphasise that the almost entirely negligible effects of the Gauss-Bonnet term on inflationary spectra in Higgs inflation-like models, as discussed in section \ref{sec:GBHstable}, may provide a natural way of realising such a theory without having to fine tune the specifics of the Gauss-Bonnet coupling to address both inflationary and post-inflationary concerns; the presently experimentally-favoured properties of standard Higgs inflation could be produced independently of the details of the Gauss-Bonnet coupling's modification of the end of inflation.

\section{Conclusions}

In this paper we have presented a tensor-scalar theory with a non-minimal coupling between the scalar field and the Ricci scalar as well as to the Gauss-Bonnet term. While this work was conceived in and motivated by the particular context of extending and generalising Higgs inflation, the analytical discussion is done in general terms, with Higgs inflation and its Gauss-Bonnet modifications just special cases identified by particular choices of coupling functions.  

To allow much of this analysis to be confidently performed in the Jordan frame, as the conformal transformation of the Gauss-Bonnet term is rather complicated, we first rediscovered the known analytical predictions for standard Higgs inflation to leading order in slow-roll, without appealing to a conformal transformation to the Einstein frame as is usually done. In doing this we utilise an additional set of slow-roll parameters pertinent to the additional terms present in the Jordan frame representation of the model. 

Upon verifying that our methods to study inflation and the primordial power spectra in a non-standard frame do not disagree with the existing literature on this topic, we proceed to apply them to the full model of interest including a non-trivial coupling to the Gauss-Bonnet combination. In doing so, we again make use of a new set of slow-roll parameters previously defined in terms of Gauss-Bonnet inflation, but modify their definition to better suit the Jordan frame and perform a leading-order slow-roll analysis to obtain expressions for the scalar and tensor spectra. 

Finally we implement a full numerical integration of the equations of motion and find that the Gauss-Bonnet term has interesting effects beyond the slow-roll approximation, such as being able to prevent oscillations of the inflaton around the minimum of its potential, instead seeming to freeze the scalar field in at a finite non-zero value. This potentially has implications for late-time cosmology and the dark energy problem (a context in which the Gauss-Bonnet term has previously been considered, see e.g. \cite{Carter:2005fu}), and in turn a link between this and early-time inflationary cosmology. We leave an analysis of the feasibility of this for future work. Furthermore, with regards to the original motivation of this work, that is, Gauss-Bonnet corrections to Higgs inflation, we find that the spectral predictions of Higgs inflation are highly stable under such corrections, suppressing any new effects from the Gauss-Bonnet coupling by several orders of magnitude due to the smallness of the field $h$ and the Hubble parameter $H$ during the observable window.

While in this work we specialised to the case of a $G \propto h^{-4}$ coupling for the sake of convenience, similar effects on the end of inflation are expected to occur for other negative power law couplings (and we have numerically confirmed this for other simple cases such as standard $m^2 \phi^2$ inflation with a $G \propto \phi^{-2}$ coupling). We further expect that due to the smallness of $H$ during Higgs inflation, corrections to the power spectra from a Gauss-Bonnet coupling will be small for any reasonable coupling function, particularly positive power law couplings in which the smallness of $h$ during the observable window will further suppress the magnitude of terms proportional to derivatives of $G$ in the equations of motion. We hence have reason to believe our results are applicable in a considerably wider context than the single choice of coupling we numerically investigated here.

\begin{acknowledgments}
The work of CvdB is supported by the Lancaster- Manchester-Sheffield Consortium for Fundamental Physics under STFC Grant No. ST/L000520/1. CL is supported by a STFC studentship. 
\end{acknowledgments}

\appendix

\section{Generalised slow-roll parameters} \label{app:GBHSRA}

In this work we have defined the usual slow-roll parameters,

\beq
\epsilon_0 = \frac{-\dot{H}}{H^2} \, , \quad \epsilon_n = \frac{\dot{\epsilon}_{n-1}}{H \epsilon_{n-1}} \, ,
\eeq

the Gauss-Bonnet flow functions, (see eg. \cite{Guo:2010jr})

\beq
\delta_0 = 4 \dot{G} H \, , \quad \delta_n = \frac{\dot{\delta}_{n-1}}{H \delta_{n-1}} \, ,
\eeq

and the Jordan frame flow functions, (see section \ref{sec:jfhinflation} and references therein)

\beq
\zeta_0 = \frac{\dot{F}}{H F} \, , \quad \zeta_n = \frac{\dot{\zeta}_{n-1}}{H \zeta_{n-1}} \, .
\eeq

The second and third of these types of slow-roll parameters are useful when considering Gauss-Bonnet inflation and Jordan frame inflation, respectively. However, when considering both the Gauss-Bonnet term and the implications of working in the Jordan frame, it is more convenient to use a different definition of the Gauss-Bonnet flow parameters. As discussed in the main text (see section \ref{sec:GBHinflation}), the more appropriate set of slow-roll parameters is,

\beq
\Delta_0 = \frac{\delta_0}{F} = \frac{4 H \dot{G}}{F}  \, , \quad \Delta_n = \frac{\dot{\Delta}_{n-1}}{H \Delta_{n-1}} \, .
\eeq

We can compare equations in terms of $\del{n}$ to ones in terms of our new $\Del{n}$ parameters by using the following relations derived from the definitions of the slow-roll parameters.

\begin{align}
\delta_0 & = F \Delta_0 \, , \\
\delta_1 & = \Delta_1 + \zeta_0 \, , \\
\delta_2 & = \frac{\Delta_1 \Delta_2 + \zeta_0 \zeta_1}{\Delta_1 + \zeta_0} \, ,
\end{align}

and so on. Expanding some common functions that appear in the equations of motion, as well as derivatives of the slow roll parameters which will be needed when calculating power spectra, we find,

\beqa
& \dot{H} =   -H^2 \eps{0} \, , \\
& \dot{G} = \Delta_0 F / 4 H \, , \\
& \dot{F} =  H F \zet{0} \, . 
\eeqa

\beqa
& \epsd{n} =  H \eps{n} \eps{n+1} \, , \\
& \deld{n} =  H \del{n} \del{n+1} \, , \\
& \zetd{n} =  H \zet{n} \zet{n+1} \, . 
\eeqa

\beqa
& \ddot{H} =  -H^3 \eps{0} \rpar{\eps{1} - 2 \eps{0}} \, , \\
& \ddot{G} = \Delta_0 F \rpar{\Delta_1 + \zeta_0 + \epsilon_0} / 4 \, \\
& \ddot{F} =  H^2 F \zet{0} \rpar{\zet{0} + \zet{1} - \eps{0}} \, . 
\eeqa

\beqa
& \epsdd{n} = H^2 \eps{n}\eps{n+1}\rpar{\eps{n+2}+\eps{n+1}-\eps{0}} \, , \\
& \deldd{n} = H^2 \del{n}\del{n+1}\rpar{\del{n+2}+\del{n+1}-\eps{0}} \, , \\
& \zetdd{n} = H^2 \zet{n}\zet{n+1}\rpar{\zet{n+2}+\zet{n+1}-\eps{0}} \, .
\eeqa

\bibliography{GBHrefs6}

\end{document}